\documentclass{aa}  
\usepackage{graphicx}
\usepackage{txfonts}
\usepackage{lscape}
\begin{document}
\title{New homogeneous iron abundances of double-mode Cepheids from high-resolution echelle spectroscopy
\thanks{Based on observations taken with the ESO 2.2-m telescope at La Silla, Chile (Proposal 073.D-0072)
and the 1.82 m telescope at David Dunlap Observatory, Canada.}
}
\author{K. Szil\'adi \inst{1}, J. Vink\'o \inst{1}, E. Poretti \inst{2}, L. Szabados \inst{3}, \and M. Kun \inst{3} }
\offprints{K. Szil\'adi}
   \institute{Department of Optics \& Quantum Electronics, University of Szeged,
              POB 406, Szeged 6701, Hungary\\
              \email{szkati@titan.physx.u-szeged.hu}
 \and 
  INAF-Osservatorio Astronomico di Brera, Via Bianchi 46, 23807 Merate, Italy
 \and 
  Konkoly Observatory of the Hungarian Academy of Sciences, POB 67, Budapest 1525, Hungary 
}

\date{Received 2007; accepted 2007}
 
\abstract
  {}
   {We define the relationship between the double-mode pulsation of Cepheids and metallicity in a more accurate way,
    determine the empirical metallicities of double-mode Cepheids from homogeneous, high-resolution spectroscopic data, 
    and study of the period-ratio -- metallicity dependence.}
   { The high S/N echelle spectra obtained with the FEROS spectrograph were analyzed using a self-developed
   IRAF script, and the iron abundances were determined by comparing with synthetic spectra assuming LTE.}
   {Accurate [Fe/H] values of 17 galactic beat Cepheids were determined. All these stars have solar or slightly subsolar
   metallicity. Their period ratio ($P_1 / P_0$) shows strong correlation with their derived [Fe/H] values. 
   The corresponding period ratio -- metallicity relation has been evaluated.}
   {}
   
\keywords{(Stars: variables:) Cepheids -- Stars: atmospheres -- Stars: general}
\authorrunning{K. Szil\'adi et al.}
\titlerunning{Iron abundances of double-mode Cepheids}
 
\maketitle

\section{Introduction}

It was realized decades ago, that the period ratio of double-mode Cepheids (i.e. classical Cepheids that 
are pulsating in two modes simultaneously, also called beat Cepheids) strongly depends on the physical
parameters of these stars, namely the mass ($M$), luminosity ($L$), mean effective temperature
($T_{\rm eff}$), and metallicity ($Z$). This can be used to give tight constraints on these physical 
parameters via comparison of observed periods and prediction of theoretical pulsational models
on the Petersen diagram ($P_1/P_0$ vs. $\log P_0$; Petersen \cite{pet73}; Morgan \& Welch
\cite{mw97}; D'Cruz, Morgan, \& B\"ohm-Vitense \cite{dcruz}). The comparison of beat Cepheids
in both Magellanic Clouds (LMC and SMC)  with their galactic counterparts revealed that the key parameter is
the metallicity. In order to get unambiguous estimates for the mass and luminosity of a given Cepheid,
its metallicity must be known from observations. 

Besides constraining pulsational and evolutionary models, metal abundances of Cepheids are useful
for studying the abundance distribution in the Milky Way (Kovtyukh et al. \cite{k05}; Lemasle et al. \cite{lem07}),
as well as in other galaxies (Beaulieu et al. \cite{b06}).  Recent theoretical calculations also indicate the importance 
of metallicity for the double-mode pulsation (Buchler \& Szab\'o \cite{bsz06}).

 Unfortunately, double-mode Cepheids are not as frequent as their single-mode counterparts, 
because radial pulsation with simultaneously excited two periods can only be maintained in a 
very limited region of the instability strip. The number of known beat Cepheids in the Milky Way galaxy
is still very small, only 23. The situation is better for LMC and SMC, where we have better statistical samples
of the beat Cepheid population, thanks to the long-term microlensing surveys. 

In this paper we present new iron abundances for an almost complete sample of known double-mode Cepheids
in the Milky Way. These data are based on analyzing of high-resolution, high signal-to-noise echelle 
spectra. The observations and the data reduction are described in Sect.~2. In Sect. 3 we give the details
of the applied method of abundance analysis. The results are presented and discussed in Sects.~4 and 5.
Finally, we draw our conclusions in Sect.~6.

\section{Observations and data reduction}

The observations were carried out with the fiber-fed high-resolution echelle spectrograph FEROS at the ESO/MPG 
2.2-m telescope at La Silla observatory, Chile (Kaufer et al., \cite{kaufer}).  FEROS has a continuous wavelength 
coverage from 3900 to 9200 \AA.  Two fibers simultaneously recorded the light from the object and the nearby sky 
background. 
The detector was a back-illuminated CCD with 2148$\times$4096 pixels of 15 $\mu$m size.
Seventeen program stars were selected as observing targets from the list of known galactic double-mode 
Cepheids accessible from the southern hemisphere (Table~1).  
No previous extensive investigation of double-mode Cepheids has ever been performed at the
resolution achieved by FEROS ($R = 48000$). 

The spectra of the program stars were collected on 3 nights between 30 May and 1 June 2004. 
The observed frames were read out in the ``normal'' mode with readout time $\sim$41.6~s, detector gain 
$\sim$3.2 $~e^-$/ADU, and readout noise $\sim$5.1 $~e^-$. The exposure times were chosen based
on primarily the brightness of the target star, but avoiding overly long exposures during which the stellar
parameters may vary significantly due to the pulsation. Calibration data, including bias, dark, flat-field, and
ThArNe lamp exposures were obtained during the standard calibration routine. 
Moreover, several daytime solar spectra were recorded with the same setup before each night. 

\begin{table}
\begin{minipage}[t]{\columnwidth}
\caption{Basic parameters of program stars. 
References: A: Antipin (\cite{ant97}); 
BP: Beltrame \& Poretti (\cite{bp}); 
PP: Pardo \& Poretti (\cite{pp});
WO: Wils \& Otero (\cite{wo04}).}
\centering
\renewcommand{\footnoterule}{}  
\begin{tabular}{lccl}
\hline\hline
Star & $P_0$ (days) & $P_1 / P_0$ & Reference \\
\hline
Y Car & 3.6398 & 0.7032 & PP \\
EY Car & 2.8760 & 0.7079 & WO \\
GZ Car & 4.1589 & 0.7054 & PP \\
TU Cas\footnote{data obtained at DDO} & 2.1393 & 0.7097 & PP \\
UZ Cen & 3.3343 & 0.7064 & PP \\
BK Cen & 3.1739 & 0.7004 & PP \\
VX Pup & 3.0118 & 0.7104 & PP \\
EW Sct & 5.8235 & 0.6986 & PP \\
V367 Sct & 6.2932 & 0.6968 & PP \\
V458 Sct & 4.8412 & 0.6993 & A \\
BQ Ser &  4.2710 & 0.7052 & PP \\
U TrA & 2.5684 & 0.7105 & PP \\
AP Vel & 3.1278 & 0.7033 & PP \\
AX Vel & 3.6732 & 0.7059 & PP \\
GSC 8691-1294 & 4.317 & 0.7035 & WO \\
GSC 8607-0608 & 4.089 & 0.7017 & WO \\
\\
V1048 Cen\footnote{1st/2nd overtone pulsator} & 0.9224 & 0.8058 & BP \\
\hline
\end{tabular}
\end{minipage}
\end{table}

In addition to the program stars observed with FEROS, two echelle spectra of the northern beat Cepheid TU Cas
were also included in the sample. These spectra were taken at the David Dunlap Observatory, Canada, with the
echelle spectrograph  mounted on the 74 inch (188 cm)  telescope in 1997. See Kiss \& Vink\'o (\cite{kv00}) for
details on the instrument and the reduction process.

The reduction of the FEROS data was done in $IRAF$, independently of the ESO/MIDAS pipeline. 
We preferred an interactive reduction, in order to check each reduction step carefully and avoid
any possible misinterpretation of the resulting spectral features. 
%

First, all frames were bias-corrected by subtracting a
master bias frame, defined as the average of several bias frames corresponding to the ``normal'' readout mode
(see above). The stellar frames were then corrected for cosmic rays with the task {\it noao.imred.crutil.cosmicrays}. 
Then, an averaged master flat-field frame was created in the same way as the master bias frame. The echelle orders 
were identified on this master flat-field frame. Since the echelle orders showed complicated double-peak
flux profiles that had a varying shape across the chip, the automatic order-finding algorithms in $IRAF$ could
not provide reliable results. Thus, we visually identified all spectral orders and manually adjusted the position
and width of each aperture centered on the particular order. This way we could reliably define and
extract the brightest 35 orders both for the object and the sky fiber, respectively. 

After this step, we independently followed two different approaches during the reduction. For the first,
we applied the {\it noao.imred.echelle.dofoe} task defined particularly for handling two-fiber-fed echelle spectra. 
This complex task performs scattered-light subtraction, order extraction based on an order-reference frame,
1D-flatfielding using the extracted flat-field orders, and wavelength calibration based on a spectral lamp frame.
The order-reference frame was the master flat-field frame described above. The extracted flat-field orders were fitted by a
smooth polynomial, keeping only the pixel-to-pixel variations, and these ``residual'' flat-fields were used in the
flat-field division. This kind of flatfielding does not correct the stellar spectra for the blaze function, 
which must be determined and removed subsequently. This was done along with the continuum normalization later.
For the wavelength calibration, a master arc frame was created by median-combining several long- and short-exposure
arc frames obtained during the calibration procedure. The reason was that the short-exposure arc spectra
had poor signal-to-noise (S/N) for many weak lines, while the long-exposure ones had too many badly saturated lines.
The combination of the arc frames was tuned to have as many weak lines with good S/N as possible with the 
smallest number of saturated strong lines. The stability of FEROS enabled us to use this single master frame for
the wavelength calibration of all object spectra obtained on the same night. The wavelength calibration was
performed within {\it dofoe} by a two-dimensional polynomial fitting to the order numbers and the 
measured pixel positions of the arc emission lines. Finally, the resulting object spectra were continuum-normalized
by fitting a smooth polynomial to the local continuum level, determined iteratively by rejecting those pixels that
deviated more than $2 \sigma$ from the local mean level. 

The second approach was based on a self-developed CL-script that includes other $IRAF$ routines
for reducing and calibrating general echelle spectra.  The reduction steps were basically the same as
described above. Specifically, order extraction was done with the {\it noao.imred.echelle.apall} task. 
Instead of fitting a global illumination pattern for the scattered light, a local background flux level was
determined and subtracted from each order during the extraction.   
The flat-field correction was done by dividing each extracted order of the object frame by the
corresponding extracted flat-field order. This way the blaze function was also  removed from the
object spectra, but the spectrum of the flat-field lamp distorted the stellar (and background) fluxes.  
However, since our purpose was to get continuum-normalized spectra, this did not cause any
problem, because the final continuum normalization easily removed the slight flux tilt caused by
the flat-field spectrum. In fact, we found that this approach produces better normalized stellar
continuum than the first approach described above, because it avoids the numerical problems
encountered in the simultaneous fitting of the strongly varying blaze function and the stellar 
continuum.  As in the first method, the wavelength calibration was based on a two-dimensional
polynomial fit to the extracted orders of the master arc frame, using the task 
{\it imred.echelle.ecidentify}. For the sky apertures, the task {\it imred.echelle.ecreidentify} was applied
based on the wavelength solution of the corresponding object aperture. As in {\it dofoe}, this accounts for
the slight zero-point shift between the wavelength solutions of the corresponding object and sky 
apertures. The removal of the sky spectra from the stellar spectra was done on the wavelength
scale, instead of pixel scale, with the task {\it imred.echelle.sarith}.  The length of each order
was then manually adjusted by trimming the noisy edges at the ends. Finally, all object spectra were
continuum-normalized using {\it noao.imred.echelle.continuum}, as described above. 

The final spectra from the two independent reductions were compared order-by-order, and very good
agreement was found. A few reduced spectra were also double-checked with those obtained by 
the imporved FEROS pipeline in MIDAS (Rainer \cite{ra03}). Very good agreement was found, 
except for a slight wavelength shift, which is due to the fact that the MIDAS pipeline contains a
barycentric Doppler-correction of the final spectra. The precision of our
wavelength calibration was tested by comparing the wavelengths of the telluric features in the
vicinity of H$\alpha$ with those from other spectra obtained with other instruments. 
Perfect agreement, within 1 pixel ($\sim$0.05\,\AA), was found.
We believe that our reduction efforts have resulted in reliable, properly wavelength-calibrated, 
continuum-normalized FEROS spectra of all our program stars.  Our calibration also supplies
a reduction approach within the IRAF environment to potential users as an
alternative to the standard MIDAS pipeline (Szil\'adi, PhD thesis, in preparation).

\section{Abundance analysis}

\begin{table*}
\caption{Atmospheric parameters and iron abundances for the program stars}
\centering                          
\begin{tabular}{l c c c c c c c c c}        
\hline\hline        
Star& Date & S/N & $T_{\rm eff}$ & $\log g$ & $\log g+a$ & $v_{\rm t}$ & $A$(Fe) & $\sigma$ & $\langle v_e \sin i \rangle$  \\
&        &        &   (K)                &   (dex)    &  (dex)         & (km s$^{-1}$) & (dex)    & (dex) &  (km s$^{-1}$)  \\  
\hline

 Y~Car&May 30 & 187 & 6000 & 2.0 & 1.92  & 4.1 & 7.45 & 0.19 & 0 \\
 &June 1 &  270 & 7000  & 2.5 & 2.48 & 4.1 & 7.43 & 0.14 & 10\\

\\
EY~Car &May 30& 142 & 6000 & 1.0 & --    & 4.2 & 7.39 & 0.24 &  --\\
&June 1& 139 & 5750  & 1.5 & --   & 4.2 & 7.24 & 0.20 & 20 \\

\\
GZ~Car&May 30& 158 & 6250 & 2.0 & 2.06  & 5.0 & 7.49 & 0.22 & 0 \\
&June 1& 125 & 6000  & 2.0 & 1.98 & 5.0 & 7.35 & 0.19 & 0 \\

\\
UZ~Cen &May 30 & 210 & 6000 & 1.0 & 2.06  & 3.8 & 7.31 & 0.21 & 4 \\
&June 1 &  168 & 5750  & 1.5 & 1.59 & 4.2 & 7.29 & 0.14& 8 \\

\\
BK~Cen &May 30 & 201 & 6000 & 2.0 & 2.08  & 4.0 & 7.65 & 0.16 & 10 \\
 &June 1 & 151 & 6000  & 2.0 & 2.17 & 4.0 & 7.48 & 0.23 & 4 \\

\\
V1048~Cen &May 30& 161 & 6250 & 2.0 & --    & 4.2 & 7.37 & 0.23 & 7 \\
&June 1&  189 & 6250  & 2.0 & --   & 4.2 & 7.25 & 0.22 & 5 \\

\\
VX~Pup&June 1& 274 & 6500  & 2.5 & 2.51 & 4.4 & 7.29 & 0.14& 10\\

\\
EW~Sct &May 30& 150 & 5750 & 1.5 & 1.57  & 3.8 & 7.48 & 0.16 & 10\\
&May 31& 197 & 6000 & 1.5  & 0.00  &  3.8 & 7.54 & 0.18& 10 \\
&June 1& 228 & 6250  & 2.0 & 2.04 & 3.6 & 7.43 & 0.20 & 10\\

\\
V367~Sct&May 30& 151 & 6000 & 1.5 & 1.33  & 4.5 & 7.54 & 0.21 & 15\\
&May 31& 146 & 6250 & 2.0  & 2.02  &  4.5 & 7.59 & 0.22 & 15 \\

\\
V458~Sct &May 30& 218 & 6250 & 2.0 & 1.98  & 4.2 & 7.55 & 0.19 & 16\\
&May 31& 183 & 6250 & 2.5  & 2.54  &  4.2 & 7.61 & 0.17 & 11 \\

\\
BQ~Ser&May 30& 133 & 5750 & 2.0 & 1.81  & 3.7 & 7.34 & 0.21 & 17\\
&May 31& 139 & 6000 & 2.0  & 1.97  &  3.9 & 7.39 & 0.27 & 16 \\
&June 1& 169 & 6000  & 2.0 & 2.03 & 3.7 & 7.36 & 0.25 & 14\\

\\
U~TrA &May 30& 300 & 6000 & 2.0 & 2.04  & 4.8 & 7.43 & 0.14 & 5\\
&June 1&  270 & 6000  & 2.0 & 2.33 & 4.8 & 7.25 & 0.18 & 0\\

\\
AP~Vel &May 30& 126 & 5750 & 1.5 & 0.00  & 3.9 & 7.51 & 0.17 & 18\\
&June 1&  154 & 6000  & 2.0 & 2.01 & 3.9 & 7.35 & 0.20 & 10\\

\\
AX~Vel&May 30& 328 & 6500 & 2.5 & 2.52  & 5.1 & 7.54 & 0.17 & 10\\
&June 1&  232 & 6250  & 2.0 & 2.17 & 5.1 & 7.26 & 0.19 & 13\\

\\
GSC~8607-0608&May 30& 169 & 5750 & 1.5 & 1.50  & 3.7 & 7.38 & 0.22 & 7\\
\\

GSC~8691-1294&May 31&  66 & 6250 & 2.5  & -- &  3.6 & 7.48 &  0.30 & 15\\
&June 1&  274 & 6000  & 2.0 & --   & 3.3 & 7.46 & 0.15 & 15 \\

\hline
\end{tabular}
\end{table*}

\begin{figure}
\centering
\includegraphics[width=8cm]{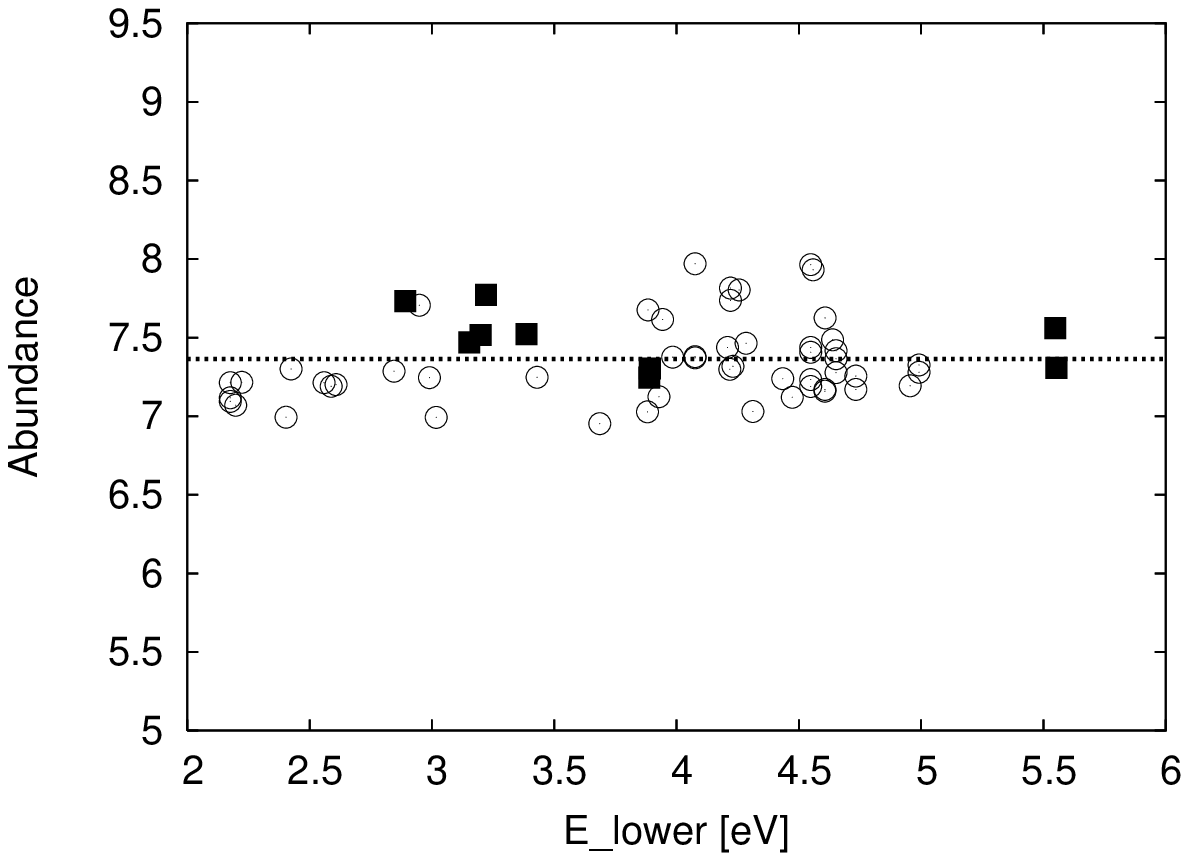}
\includegraphics[width=8cm]{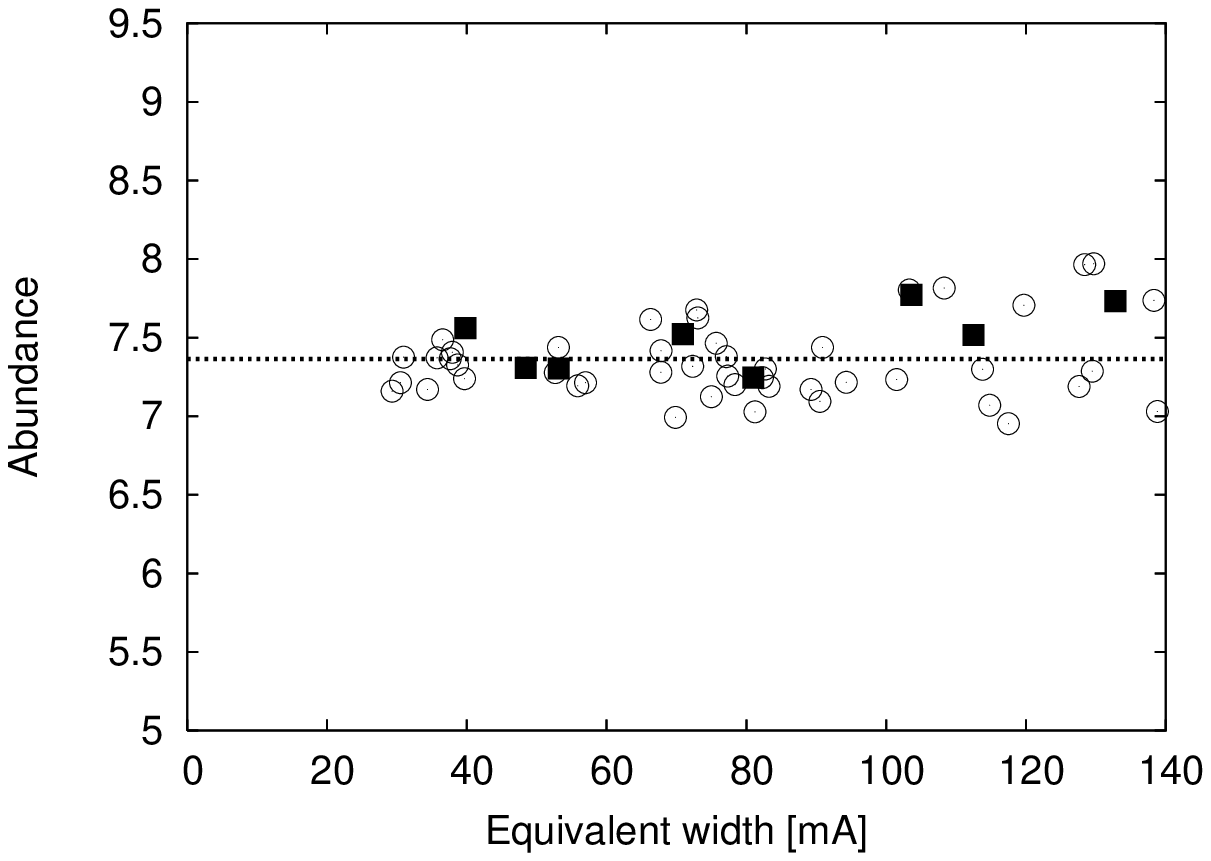}
\caption{Fe abundances for BQ Ser with optimal atmospheric parameters. 
The parameters were set as $T_{\rm eff} = 6000$ K, $\log g = 2.0$, 
$v_{\rm t} = 3.7$ km\,s$^{-1}$. Open circles denote Fe~I lines, filled squares are
Fe~II lines. Top panel: iron abundance vs. excitation potential; bottom panel: abundance vs. EW.}
\end{figure}

\begin{figure}
\centering
\includegraphics[width=8cm]{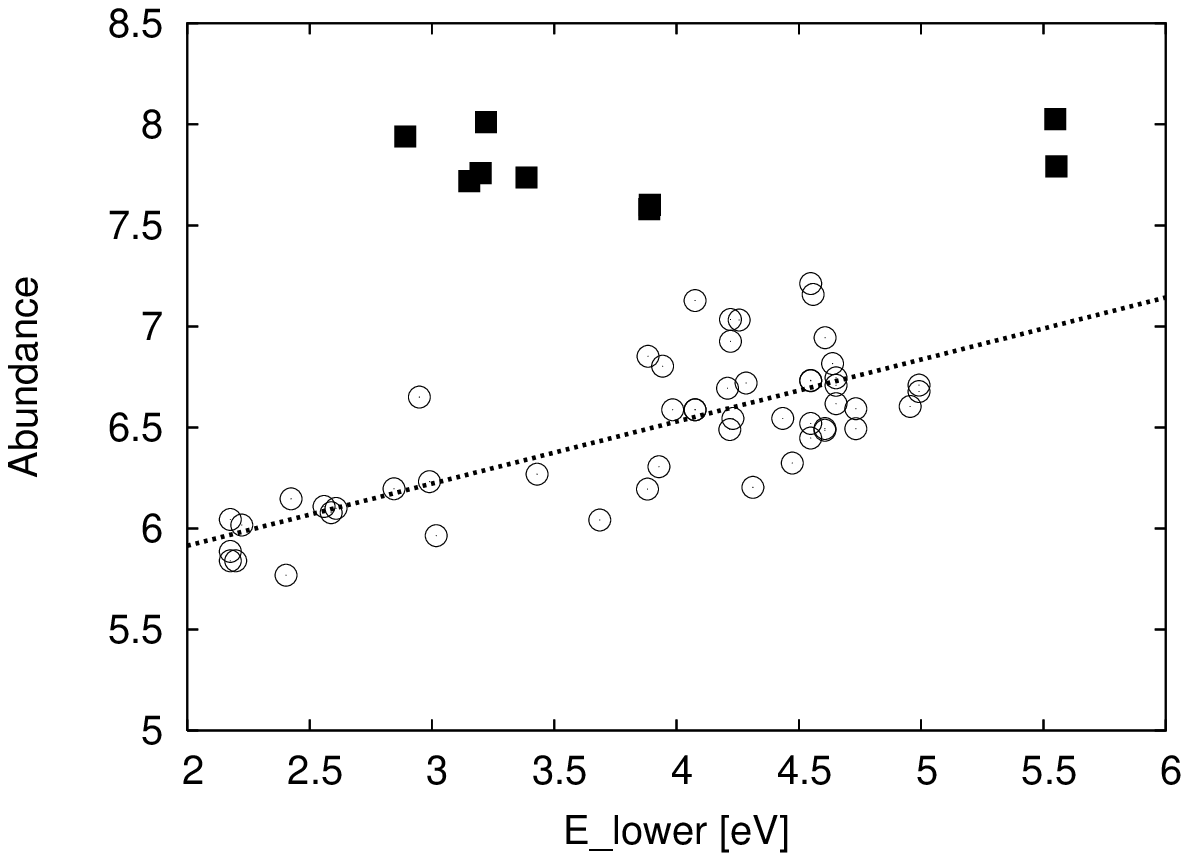}
\includegraphics[width=8cm]{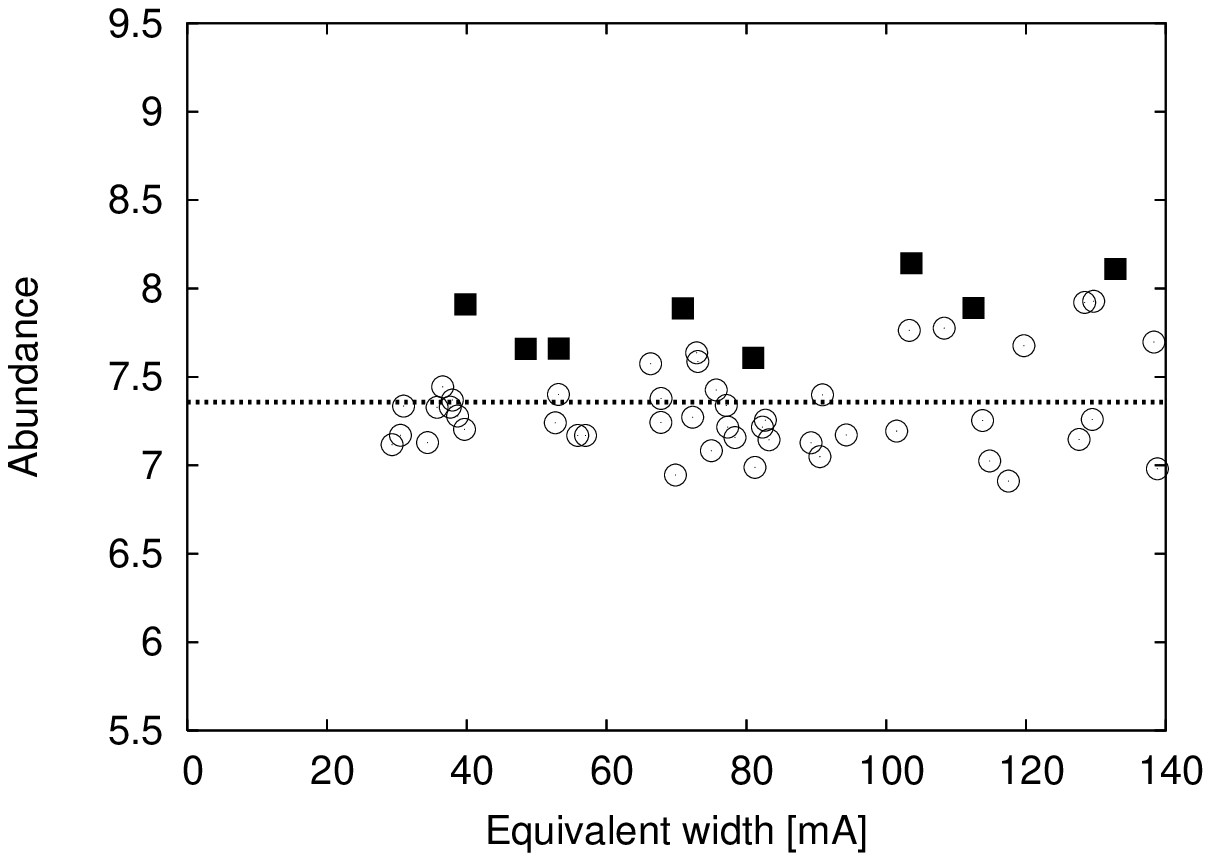}
\includegraphics[width=8cm]{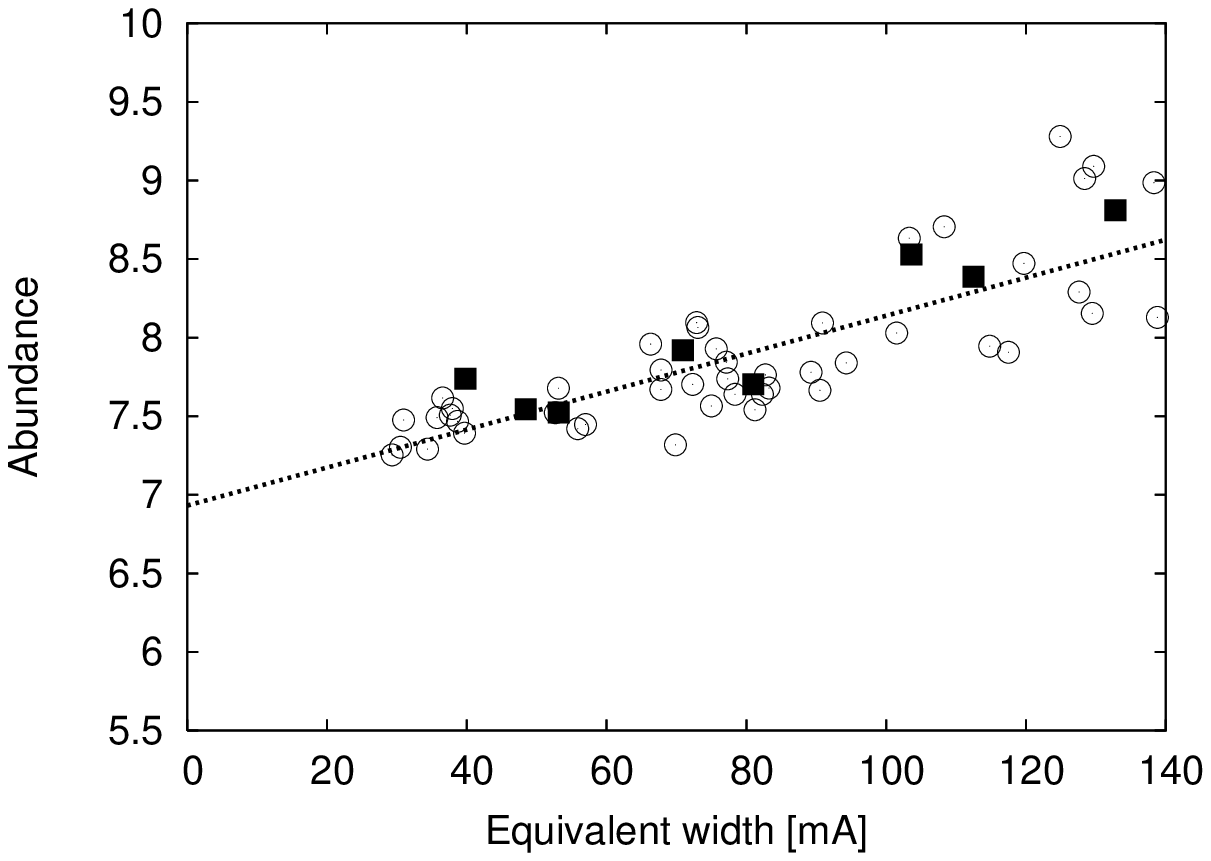}
\caption{The sensitivity of the calculated line abundances to the atmospheric parameters
in the case of BQ Ser. The meaning of the symbols is the same as in Fig.~1. 
Top panel: the effect of setting $T_{\rm eff} = 5500$ K; 
middle panel: setting incorrect gravity ($\log g = 3.0$); 
bottom panel: incorrect microturbulence ($v_{\rm t} = 1.5$ km\,s$^{-1}$).}
\end{figure}

\begin{figure}
\centering
\includegraphics[width=8cm]{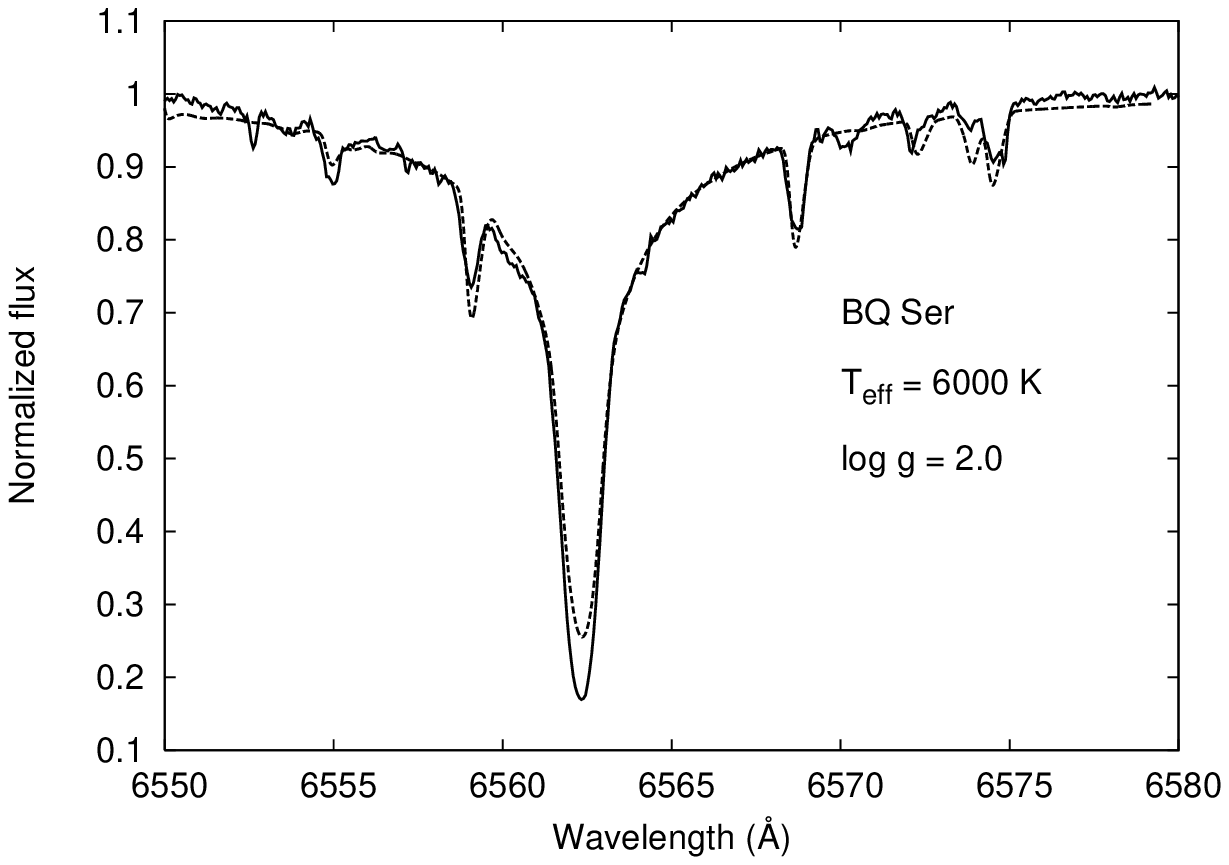}
\caption{The fitting of the H$\alpha$ profile of BQ Ser. The observed spectrum is shown by continuous
curve, while the model spectrum is plotted as a dashed curve. The model shown here was computed assuming 
$T_{\rm eff} = 6000$ K and $\log g = 2.0$.  The fit was restricted to the wings of
H$\alpha$, since the core cannot be modeled under the LTE assumption.}
\end{figure}

In order to determine the iron abundance of the program stars, first, a list of weak, unblended
Fe~I and Fe~II lines was assembled based on the information (wavelength, oscillator strength,
and excitation potential) found in the extensive line list of Kovtyukh \& Andrievsky (\cite{ka99}). 
Each line profile was visually examined and blended ones were rejected. This resulted in 
77 Fe~I and 18 Fe~II lines. 
Then, the equivalent widths (EWs) of these lines were calculated from the observed spectra using 
the $IRAF$ task {\it splot}. First, the spectrum was Doppler-corrected (with the task {\it dopcor})
to bring the lines as close to their laboratory wavelength as possible. The input radial velocity
was calculated via cross-correlating the stellar spectrum with a daytime solar spectrum, which has
approximately zero radial velocity with respect to the observer. The 20th echelle order containing the
MgI `b' triplet at $5184$ \AA~, which is often used for radial velocity studies of late-type stars, 
was selected for this purpose. The cross-correlation function (CCF) was computed with the
{\it fxcor} task. Having a rest-frame corrected spectrum, the EWs of the selected lines
were derived with the task {\it splot} by fitting Voigt-profiles, allowing slight wavelength shifts.
Those lines that showed shifts greater than $\pm 0.2$ \AA~ from their laboratory position were omitted
from the later analysis. 

For each line the iron abundances were determined with the 
SPECTRUM\footnote{http://www.phys.appstate.edu/spectrum/spectrum.html} 
spectral synthesis code by R.O. Gray. SPECTRUM solves the radiative transfer
equation in a plane-parallel model atmosphere assuming LTE. We used Kurucz
model atmospheres with $T_{\rm eff} = 5000$-$7000$ K, $\log g = 0$-$2.5$
dex as appropriate for Cepheids, and a solar abundance pattern. 
After selecting a particular atmosphere (specifying $T_{\rm eff}$ and $\log g$)
and microturbulence $v_{\rm t}$, the LTE abundances 
$ A({\rm Fe})~=~log(N_{\rm Fe}/N_{\rm H})+12$,
where $N_{\rm Fe}$ and $N_{\rm H}$ is the number density of iron and
hydrogen, respectively, were calculated with the 
auxiliary code BLACKWEL for each input line. To minimize the
number of lines that are affected by NLTE-effects, only weak lines with
EW $< 0.15$ \AA~ were used. 

The atmospheric parameters were determined entirely from the observed
EWs in the standard way (e.g. Fry \& Carney \cite{fc97}). The effective
temperature is selected by requiring that the calculated abundances do not
depend on the line excitation potential. Similarly, the surface gravity is chosen
when the calculated abundances are the same for both Fe~I and Fe~II lines
(i.e. the abundances do not depend on the ionization state). 
Both $T_{\rm eff}$ and $\log g$ were further constrained by fitting 
the profiles of H$\alpha$ with model spectra calculated with ATLAS9/SYNTHE. 
Finally, the microturbulent velocity is set when the line abundances are independent of the line EWs. 

 Figure~1 illustrates the correct choice of these parameters
in the case of BQ Ser. The final iron abundance (indicated by the horizontal line) 
was calculated as the average of all individual abundances including neutral and ionized 
Fe lines.  The top, middle, and bottom panels of Fig.~2 show the effect of an incorrect temperature,
gravity, and $v_{\rm t}$, respectively. Decreasing either $T_{\rm eff}$ from 6000 K to 5500 K
(top panel)  or the microturbulence from  3.7 to 1.5 km s$^{-1}$ (bottom panel), a strong 
dependence of the Fe abundance on the excitation potential is introduced. 
A change in  $\log g$ from 2.0 to 3.0 (middle panel) causes a systematic difference between 
the abundances calculated from Fe~I and Fe~II lines.

One example of the H$\alpha$ profile fitting is plotted in Fig.~3.  Due to its temperature and
pressure sensitivity, the H$\alpha$ line is a good indicator of the photospheric 
temperature and gravity, even though the line core is more affected by NLTE effects. Thus, 
the fitting is restricted only to the line wings. Since the abundance data are only weakly
sensitive to the gravity (see Fig.~2), the H$\alpha$ fitting provided important constraints
for this parameter. 

To test the abundance scale of BLACKWEL, as well as the
accuracy of our EW measurements, the daytime solar spectra were 
also analyzed using the same line list. The Kurucz solar model 
atmosphere and its parameters 
($T_{\rm eff} = 5777$ K, $\log g = 4.4377$, $v_{\rm t} = 1.5$ km\,s$^{-1}$)
were used and the resulting line abundances averaged. 
The results were $A_{\odot}({\rm Fe}) ~=~ 7.44 \pm 0.18$,  $7.52 \pm 0.18$, 
and $7.51 \pm 0.17$ for 30 May, 31 May, and 1 June, respectively, 
where the uncertainties were estimated as the standard
deviation of the line abundances around the mean.
It is visible that for the spectrum obtained on the first day (30 May)
the calculated abundance is smaller than those from the two other days,
but they agree well within $1 \sigma$. Thus, we adopted their
unweighted average $A_{\odot}({\rm Fe}) ~=~ 7.49 \pm 0.18$ as the
final solar abundance. This value is in remarkable agreement with the
recent solar iron abundance of $7.45 \pm 0.05$  
(e.g. Asplund et al. \cite{a05}), and $A_\odot ({\rm Fe}) ~=~ 7.49$ was adopted
as the observed zero point for determining [Fe/H] from
the FEROS spectra. 

\section{Results}

The iron abundances, together with the corresponding atmospheric parameters, are collected in
Table~2. The uncertainties of $T_{\rm eff}$ and $\log g$ are $\pm 250$ K and $\pm 0.25$ dex, respectively,
due the resolution of the applied grid of model atmospheres. The estimated uncertainty of the
microturbulent velocities is $\pm 0.5$ km\,s$^{-1}$. The errors of the abundances are
purely statistical, being the rms standard deviation of the individual line abundances around
the mean value (see the previous section). 

It is important to note that the inferred surface gravities are actually the effective gravities
containing the acceleration term due to the pulsation
\begin{equation}
g_{\rm eff} ~=~ {{G M} \over {R^2(t)}} ~-~ p {{d v_{\rm rad}} \over {dt}}
\end{equation}
where $M$ is the stellar mass, $R(t)$ the instantaneous radius of the photosphere,
$p$ the projection factor to convert the radial velocities into the
pulsational velocities (we used $p = 1.38$ from the period-dependent formula 
by Gieren et al. \cite{g99}), 
and  $v_{\rm rad}$ the radial velocity at the moment of the observation. 

To get information on the true gravities of the program stars, we estimated the 
acceleration terms from the derivative of the radial velocity curves. 
The radial velocity data were collected from  Stobie \& Balona (\cite{sb}),
Gorynya et al. (\cite{gor92}, \cite{gor96}, \cite{gor98}), Bersier et al. (\cite{ber}), 
Antipin et al. (\cite{ant}), and Petterson et al. (\cite{pet}). 
For each Cepheid, 
the radial velocities were fitted by a Fourier-polynomial including the
 $f_0$, $f_1$, $f_0 + f_1$, $f_1-f_0$, and $2f_0 + f_1$ frequency components. 
 The frequencies were computed from the periods listed in Table~1. Because these are
determined from observations distributed on a long time baseline, they can be
applied to derive the phases of the new observations unambiguously.
Then, the analytic derivative of the Fourier-polynomial was computed at the phase of the given 
spectroscopic observation to get the acceleration term in Eq.~1. Table~2 also contains the true
gravities, i.e. the observed (effective) gravities corrected for the acceleration. 

The chemical abundances should not depend on time, so the data in Table~2 can also be
used to test the stability and the possible systematic errors of our abundance analysis.
Generally, it is visible that the calculated abundances do agree within their errors.
However, the spectra from day~3 (June 1) tend to give slightly lower abundances 
(but within $1 \sigma$) than those from day~1 (May 30). No such tendency can be 
seen in the atmospheric parameters. We attribute this slight inconsistency to an
instrumental calibration problem, probably associated with the scattered light + sky 
background removal that may alter the line depths of continuum-normalized spectra. 
Nevertheless, since the differences are not significant, we define the final iron abundances
(Table~3) as the average of the values given in Table~2. 
Their final estimated uncertainties reflect
their individual errors, plus the differences between the data belonging to different nights. 
In Table~3 we also list the new [Fe/H] values calculated from 
\begin{equation}
{\rm [Fe/H]} ~=~ A_{\rm Ceph}({\rm Fe}) ~-~ A_\odot ({\rm Fe})
\end{equation}
where $A_\odot ({\rm Fe}) = 7.49$ was applied (see the previous section).
We note that 14 out of 17 double-mode Cepheids show slightly negative [Fe/H] values,
though while considering their errorbars, the deviation from solar
abundance is marginal. 

\begin{table*}
\caption{Final (averaged) abundances and uncertainties with [Fe/H]
data from previous studies. For the latter data, uncertainties  
(where available) are given in parentheses.  }
\begin{center}
\begin{tabular}{lrccccc}
\hline\hline
\multicolumn{1}{c}{Star} & \multicolumn{1}{c}{[Fe/H]} & $\sigma$ & [Fe/H]$_1$ & [Fe/H]$_2$ & [Fe/H]$_3$ & [Fe/H]$_4$ \\
    & \multicolumn{1}{c}{(dex)} & \multicolumn{1}{c}{(dex)} & (dex) & (dex) & (dex) & (dex) \\
\hline
Y~Car  &    $-$0.05  & 0.16  &  $-$0.33 (0.35)   &   --    &  --   &   --   \\
EY~Car  &    $-$0.17 & 0.22  &  --      &   --    &  --   &   --   \\
GZ~Car  &    $-$0.07  & 0.20  &  $-$0.36 (0.33)   &   $-$0.08 &  --   &   --   \\ 
UZ~Cen  &    $-$0.19  & 0.17  &  $-$0.30 (0.32)  &   $-$0.09 &  --   &   --   \\
BK~Cen  &     0.07 & 0.19  &  $-$0.03 (0.30)  &   --    &  --   &   --   \\
VX~Pup &    $-$0.20  & 0.14  &  $-$0.39 (0.14)  &   $-$0.17 & $-$0.13 &   $-$0.15 (0.12)\\ 
EW~Sct  &    $-$0.01  & 0.18  &  $-$0.08 (0.09)  &   --    & +0.03 &   --   \\ 
V367~Sct &     0.075 & 0.21  &  +0.26 (0.30)  &   --    & $-$0.01 &   --   \\ 
V458~Sct &     0.09  & 0.18  &  --      &   --    &  --   &   --   \\ 
BQ~Ser &    $-$0.13  & 0.24  &  $-$0.36 (0.20)   &   $-$0.08 & $-$0.05 &   --   \\
U~TrA &    $-$0.15  & 0.16  &  $-$0.31 (0.34)  &   --    &  --   &   --   \\
AP~Vel &    $-$0.06 & 0.18  &  $-$0.18 (0.35)  &   $-$0.045&  --   &   --   \\
AX~Vel &    $-$0.09 & 0.18  &  $-$0.43 (0.29)  &   $-$0.12 &  --   &   $-$0.15 (0.07) \\ 
GSC~8607-0608 &    $-$0.11  & 0.22  &  --      &   --    &  --   &   --   \\
GSC~8691-1294 &    $-$0.015 & 0.22  &  --      &   --    &  --   &   --  \\ 
TU Cas & $-$0.16 & 0.24 & $-$0.43 (0.31) & $-$0.05 & 0.02 & -- \\
\\
V1048~Cen  &    $-$0.18  & 0.22  &  --      &   --    &  --   &   --   \\
\hline
\end{tabular}
\end{center}
References to [Fe/H] values: 
1: Andrievsky et al. (\cite{a94}); 2: D'Cruz et al. (\cite{dcruz});
3: Andrievsky et al. (\cite{a02}); 4: Lemasle et al. (\cite{lem07}).
\end{table*}

\begin{figure*}
\centering
\includegraphics[width=16cm]{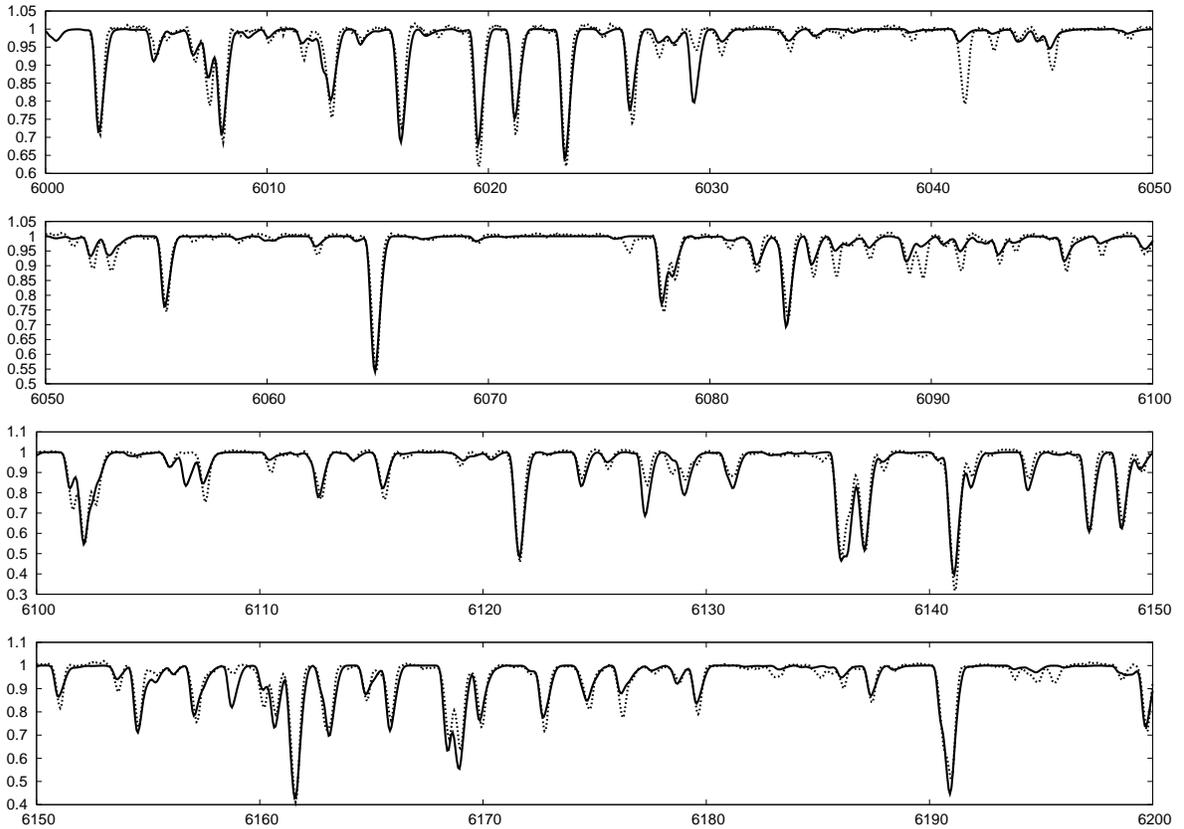}
\caption{Comparison of the observed spectrum of EW Scuti (dotted curve) with a model spectrum
computed by ATLAS9/SYNTHE (solid curve) with the parameters in Table~2. The solar abundance
pattern is scaled to the calculated abundance of EW Sct. Pulsation, as well as
the instrumental broadening, has been taken into account (see text).}
\end{figure*}

We checked our new iron abundances further by computing synthesized
spectra with ATLAS9 using Kurucz models with solar abundance
pattern scaled to the abundances in Table~2. Note that the solar iron
abundance applied in the Kurucz models corresponds to 7.63 on our scale,
so the ``abundance scale'' parameter was set to $10^{[A({\rm Fe}) - 7.63]}$, where
$A({\rm Fe})$ is the iron abundance of the Cepheid. Moreover, we enabled
the pulsation correction in VFIELD with the appropriate velocity
value determined previously. The final spectra are convolved with
the FEROS instrumental profile derived from fitting Gaussians
to the spectral lamp emission lines. In Fig.~4 a synthesized spectrum 
is compared with an observed spectrum of EW Scuti.  The agreement
is satisfactory, although a few weak lines are not fitted well.
The cause of this is that the $\log gf$ values in the Kurucz line list were
not adjusted to the values used by us in the abundance analysis (see above), 
which makes the strength of
these lines different from those obtained with the SPECTRUM code
using the same abundance. 
This test also illustrates that the iron line strengths and 
abundances computed by two independent codes, SPECTRUM and
ATLAS9, are generally consistent.

The new [Fe/H] data in Table~3 allow us to update the observed 
period ratio -- the metallicity relation first pointed out by Andrievsky et al. 
(\cite{a93}, \cite{a94})  in graphical form. From Tables~1 and~3, the 
following empirical relation has been found:
\begin{equation}
\begin{array}{lclclcl}
P_1 / P_0  &  =  & -0.0143 & \log P_0 & - 0.0265 & {\rm [Fe/H]} &  + 0.7101 \\
                 &  & \pm 0.0025 &         & \pm 0.0044    &              & \pm 0.0014.   \\
\end{array} 
\end{equation}
The period ratio, corrected for the weak dependence on the fundamental
period, is plotted against [Fe/H] in Fig.~5.

\begin{figure}
\centering
\includegraphics[width=8cm]{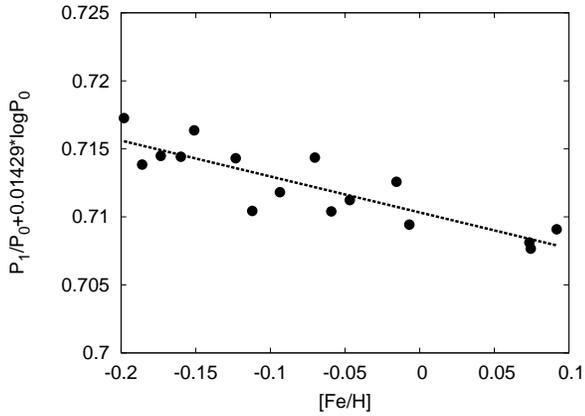}
\caption{The period ratio -- metal abundance relation. The parameters of the dotted line
are given in Eq. 3. See text for the correction term on the vertical axis.}
\end{figure} 

\subsection{The effect of rotation and pulsation on the derived abundances}

Rotation and/or pulsation alter the line profiles, thus, their effect on the 
abundance analysis needs to be investigated. These macroscopic motions 
convolve the local line profile over the visible stellar surface, therefore 
the observed (integrated) line broadens and becomes shallower. 
As a first approximation, the total EW of the observed line
does not change because of the conservation of the total absorbed flux.
However, there are second-order effects, connected with e.g. the measurement 
of the EWs from the observed spectra with finite resolution and S/N, which may
introduce some systematic uncertainties in the EWs of broadened lines, and
these effects should be evaluated. We show below that in the case of 
our program stars neither rotation nor pulsation affects the results of the
abundance analysis significantly.

Cepheids are slow rotators, similar to other
supergiants as expected from stellar evolution. The typical value of
$\langle v_e \sin i \rangle$ is $\sim 10$ km s$^{-1}$ (De Medeiros et al. \cite{demed}).
There seems to be a correlation between the rotation and pulsation period: 
Cepheids with longer pulsation periods are slower rotators
(Nardetto et al. \cite{nard}). However, Takeda et al. (\cite{tak}) found
$\langle v_e \sin i \rangle = 0$ for the short-period Cepheid SU Cas, which illustrates
that even the shortest period (hence the smallest) Cepheids may be very slow
rotators. Buchler and Szab\'o (\cite{bsz06})
studied the effect of rotation on the period ratio of beat Cepheids from theoretical point of view, and find that the period
ratio and its dependence on heavy element abundance is practically insensitive to the
rotation (up to 20 km s$^{-1}$).

To estimate the projected rotational velocities of our
program stars, we determined the FWHM
of a few weak, unblended lines (Fe~I $\lambda \lambda$6056, 6096, 6157, 6165 \AA) and
compared it with those of ATLAS9 model spectra ($T_{\rm eff} = 5750$ K,
$\log g = 1.5$, and $v_t = 4.2$ km s$^{-1}$) rotated with 0, 5, 10, 15, and 20 
km s$^{-1}$.  We checked that variations in the atmospheric parameters 
up to 250 K, 0.5 dex, and 0.5 km s$^{-1}$ in the temperature, gravity, and 
microturbulence, respectively, do not alter the FWHMs of the rotationally broadened lines
significantly. Thus, instead of computing detailed models for all program stars, 
we used only this sequence of theoretical model spectra as templates for estimating the rotational
broadening. As for the earlier results mentioned above, all program stars 
were found to be slow rotators with $\langle v_e \sin i \rangle ~<~ 20$ km s$^{-1}$ (see Table 2).

Since pulsation causes an additional broadening, using a non-pulsating
template spectrum causes a systematic overestimation of the rotational velocities.
Because of this, our estimated $\langle v_e \sin i \rangle$ values are actually upper limits to
the true projected rotational velocities.  
To test the influence of pulsation on the model FWHMs, we added
$+20$ km s$^{-1}$ pulsational velocity (typical of our short-period Cepheids)
to the model sequence and re-computed the FWHMs of the same lines. It was
found that this pulsational velocity causes $\sim 5$ km s$^{-1}$ overestimation
of the $\langle v_e \sin i \rangle$, which is about the same as the uncertainty of the estimated
rotational velocities ($\sim 3$ km s$^{-1}$). Note that the FWHM of the 
instrumental broadening profile of FEROS is $\sim 0.15$ \AA, which 
corresponds to $\sim 7$ km s$^{-1}$ rotational velocity at 6000 \AA. 
Thus, $\langle v_e \sin i \rangle$ values lower than this level are not resolved well from
our spectra.

Finally, the effect of rotation and pulsation on the model EWs was found to be
1 \% for $v_{\rm rot} = 20$ km s$^{-1}$ and less than 0.1 \% for 
$v_{\rm pul} = 20$ km s$^{-1}$. These numbers illustrate that the effect
of rotation on the line EWs and the abundance analysis presented in the previous 
sections is negligible.
  
Pulsation does not change the line EWs, but it does change the atmospheric parameters 
($T_{\rm eff}$, $\log g$) along the pulsational cycle. However, the values reported
in Table~2 show that the variation in the derived abundances is typically less than 
the errors of the individual abundances, caused by mainly
the uncertainties in the atomic data and the model atmospheres. Therefore, we conclude
that, although for pulsating stars the determination of the atmospheric parameters
is more complex than in the case of static stars, the analysis resulted in consistent
iron abundances for different pulsational phases. Thus, the abundances for those stars
that were observed only in a single phase are also probably not affected by the
pulsational motion.

\section{Discussion}

In this paragraph we compare the spectroscopically determined physical parameters
($T_{\rm eff}$, $\log g$, [Fe/H]) of the program stars with results from previous
observational studies and predictions from pulsation theory. In addition, we
give an estimate of the color excess $E(B-V)$ of the program stars from
the spectroscopic effective temperatures.

\subsection{Comparison with previous {\rm [Fe/H]} determinations}

Among the stars listed in Table~3, four targets have [Fe/H] values
determined in this paper for the first time. However, for the majority of the program 
stars  more than one 
value exists for iron abundance published previously, thus, it is important to
compare our results with those of earlier studies. 
The results collected from the literature are also summarized in Table~3. 
In Fig. 6, the literature [Fe/H] values are plotted
with respect to the ones determined in this paper.
The uncertainties of our new [Fe/H] data are between $0.1$ - $0.2$ dex (Table~3).

There seems to be a systematic difference of about
$1 \sigma$ between our data and those of 
Andrievsky et al. (\cite{a94}) in the sense that
the latter show generally smaller iron abundance for
the majority of our beat Cepheid sample. 
However, the [Fe/H] zero point of Andrievsky et al.
(\cite{a94}), $7.51$, agrees quite well with ours ($7.49$); 
thus, the cause of this discrepancy is probably 
the use of different input atomic data and Cepheid line EWs.
Also, the spectral resolution of their data is lower than ours, so
the corresponding errors  are higher ($\sim 0.3$ dex) 
than the uncertainty of the new data presented in this paper. 
On the other hand, the more recent [Fe/H] determinations
by Andrievsky et al. (\cite{a02}) agree much better with our data,  although only 5 stars are common in their
more extended observational sample with ours. We believe
that the agreement of the more recent (independent) data increases the 
probability that these are closer to the true [Fe/H] values of these
stars than the earlier ones.   Although Andrievsky et al. (\cite{a02})
do not specify individual error bars for their [Fe/H] values, they 
note that the average uncertainties of their abundances are
between $0.05$ - $0.2$ dex, which is comparable to the error bars
of our data.

The agreement is also good (within $1 \sigma$) with the data
by D'Cruz et al. (\cite{dcruz}), who predict the [Fe/H] values
from theoretical pulsational models (and also point out the
discrepancy between their results and those of Andrievsky
et al. (\cite{a94}).  Unfortunately, they also do not assign
errors to their theoretical [Fe/H] values except a short note
that ``they should be quite accurate, provided the theoretical
opacities and models are sufficiently accurate". Indeed, it is
seen from Table~3 that they agree with 
our new spectroscopic abundances within the errors.

Finally, only two stars are common in the sample of 
Lemasle et al. (\cite{lem07}) and ours.  For both stars the 
results agree well with our ones and also with those of 
Andrievsky et al. (\cite{a02}). 

\subsection{The metallicity of the 1O/2O pulsator V1048 Cen}
 The case of V1048 Cen ($\equiv$HD~304373) deserves a particular discussion. 
Together with CO Aur, they are the only 1O/2O pulsators known in
the Milky Way (Beltrame \& Poretti \cite{bp}). 
Its position in the Petersen diagram  is very similar to that of other 
1O/2O pulsators in the Large Magellanic Cloud.
This similarity is striking, considering that the F/1O pulsators 
in the Milky Way and in the Magellanic Clouds
describe well-separated sequences in the Petersen diagrams
(see Fig.~1 in Poretti \& Beltrame \cite{pb04}).
It could be argued  that V1048 Cen (and  then CO Aur, too)
could have a particular metallicity, different from those of other double-mode Cepheids.
However, our data (see Table 3) demonstrate  that the
metallicity of V1048 Cen is very similar to those of galactic F/1O pulsators.
Therefore,  it still remains unexplained why the two classes of double-mode
pulsators behave differently in different galaxies.

\subsection{Period--gravity relation} 

The linear pulsation theory predicts a simple relation 
between the period of the fundamental mode $P_0$
and surface gravity, namely,
\begin{equation}
\log g ~=~ 2.62 ~-~ 1.21 \log P_0
\end{equation}
(Kov\'acs, \cite{kg00}).  This can be inferred 
from a simple pulsation equation and blackbody 
assumption.  Using the spectroscopic gravities
listed in Table~2, it is possible to test this simple
theoretical  relation with observations. 

Since $\log g$ in Eq. 4 refers to the mean gravity,
averaged over the pulsational cycle, the gravities 
corrected for the acceleration term (Table~2)  were 
averaged. Only those stars that had more than one
spectra were considered, although the average of two
gravities obtained at random pulsational phases may
be quite off from the proper mean value. Those stars were also omitted
for which there were not enough velocity data to
compute the correction for acceleration (see Sect.~3).
This resulted in 10 stars in the 
subsample. Their averaged (corrected) gravities are
plotted in Fig.~7 as a function of $\log P_0$. (The error bars
are estimated as $\pm 0.25$ dex, although this may be
optimistic due to the small number of averaged data.) 
The linear least-square fit to the data has the parameters
\begin{equation}
\begin{array}{lcrcrc}
\log g &~=~& 2.76 & - & 1.20 & \log P_0,\\
          &       & \pm 0.33 & & \pm 0.54 &,\\
\end{array}
\end{equation}
while the theoretical relation is from Eq. 4.
It is seen that the observational data support the
theoretical relation nicely. The period dependence is fully
recovered with only a slight zero-point difference. 
This gives further confidence to the reality of the 
calculated atmospheric parameters and abundaces in
Table~2.  A similar diagram is presented by Andrievsky
et al. (\cite{a02}) for a larger sample of Cepheids, but they
plotted the uncorrected gravities (containing the acceleration term)
against $\log P_0$.

\subsection{The period ratio -- metallicity relation}

It is striking that, based on high-resolution spectroscopic observations,
the period ratios of the majority of known galactic double-mode Cepheids depend
mostly on their metal abundance and (weakly) on the pulsational period.
Since the simultaneous excitation of two pulsational modes (i.e. double-mode
pulsation), and particularly the period ratio itself,  strongly depends on other 
stellar properties (mass, mean effective temperature, luminosity) as well,
the dominance of the metallicity dependence suggests that the observed galactic
beat Cepheids should have nearly the same mass and effective temperature.
Indeed, the theoretical masses of several of these stars inferred by 
D'Cruz et al. (\cite{dcruz}) show a very low scatter of $\sim 0.3~M_\odot$ around
the mean value of $4.2~M_\odot$.  

We have also found an agreement  between the observed period ratios
in the Magellanic Clouds and the extrapolation of the galactic relation toward 
lower [Fe/H] values. This may mean that the slope of galactic relation is not strongly 
biased by the small number of galactic Cepheids analyzed here.  
A more thorough study 
of the metallicity distribution of Magellanic Cloud beat Cepheids will be the topic
of a subsequent paper.

\subsection{Determining the color excess from spectroscopic temperatures}
 The effective temperatures obtained in the manner described above can be used for determining the color excess 
of the target Cepheids. The unreddened $B-V$ color index corresponding to the given $T_{\rm eff}$ value was 
taken from Table~15.7 in Cox (\cite{cox00}) by interpolating the relevant data. With this lack of simultaneous photometry, 
the instantaneous `observed' $B-V$ color index was determined from Berdnikov's (\cite{berd06}) extensive data 
set of Cepheids which contains multicolor photoelectric observations for 13 stars discussed in this paper 
(among several hundred other Cepheids). When predicting the color index for the moment of spectroscopic 
observations, the frequencies corresponding to the periods listed in Table~1 were used, and three harmonics of 
the fundamental frequency and the sum and the difference of the frequencies of the two excited modes were taken into account. 
Thus a pair of reddened and unreddened $B-V$ values was generated for each moment 
of spectral observation. Because the color excess should not vary during a pulsational cycle, the color excesses 
were averaged for the individual beat Cepheids, and these $E(B-V)$ values are listed in Table~4. The comparison 
of the color excess data with those obtained previously by others and our own independently derived other values 
(from infrared photometry) will be discussed in a subsequent paper. The accuracy of our color excess data can be judged 
from the physically unrealistic, slightly negative $E(B-V)$ values obtained for the brightest (i.e. nearest) stars in this sample. 
Both UZ~Cen and U~TrA can be slightly reddened, therefore the error of the color excess derived here may be 
about 0.05-0.06 mag.

\begin{table}  
\caption{The $E(B-V)$ color excess of the program stars with available $BV$ photometry} 
\centering
\begin{tabular}{lr}
\hline\hline
\multicolumn{1}{c}{Star} & \multicolumn{1}{c}{$E(B-V)$} \\
\hline
Y~Car  &  0.175  \\
EY~Car  & 0.135  \\
GZ~Car  & 0.423  \\ 
UZ~Cen  &  $-$0.023 \\
BK~Cen  &  0.163 \\
VX~Pup &   0.241 \\ 
EW~Sct  &  1.089 \\ 
V367~Sct & 1.273  \\ 
V458~Sct &  0.636  \\ 
BQ~Ser &  0.854  \\
U~TrA &  $-$0.043  \\
AP~Vel &   0.371 \\
AX~Vel &   0.261 \\ 
\hline
\end{tabular}
\end{table}


\begin{figure}
\centering
\includegraphics[width=8cm]{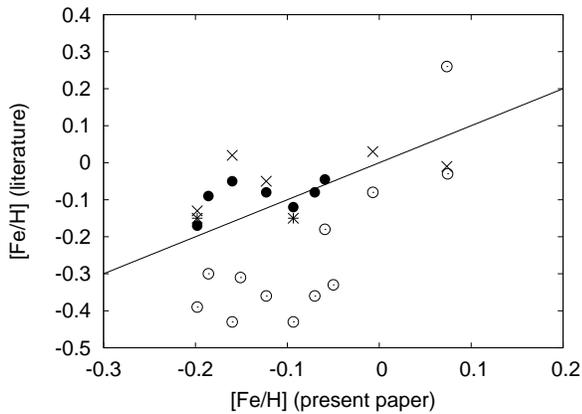}
\caption{Comparison of the [Fe/H] data of this paper with values from the literature (see 
Table~3 for tabulated data).  The meaning of the symbols is as follows:
filled circles: D'Cruz et al. (\cite{dcruz}); open circles: Andrievsky et al. (\cite{a94}); 
crosses: Andrievsky et al. (\cite{a02}); asterisks: Lemasle et al. (\cite{lem07}).
The reason for the discrepancy between our values and those by Andrievsky et al. (\cite{a94})
is described in the text.}
\end{figure}

\begin{figure}
\centering
\includegraphics[width=8cm]{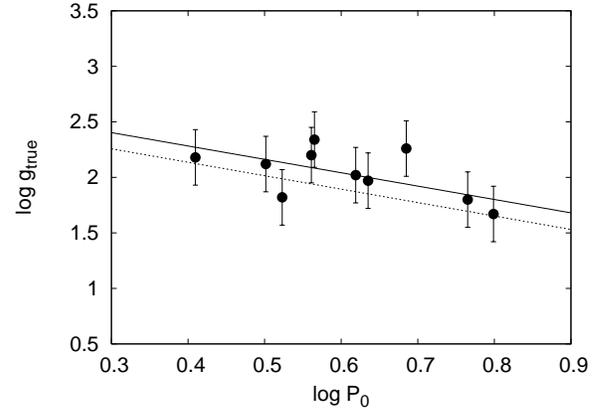}
\caption{Period -- surface gravity relation from observations
and theory. The dots are the averaged, corrected surface
gravities of the program stars (see text). The dotted line is
the relation predicted by simple pulsation theory (Eq.~4),
while the solid line is the result of a least-square fit (Eq.~5). 
The theoretical relation is nicely recovered from the observations.}
\end{figure}

\section{Conclusions}

New, accurate, homogeneous iron abundances have been derived for a large number of
galactic double-mode Cepheids (17 out of 23) 
from unprecedented high-resolution, high S/N echelle spectra. For 4 of the program stars these are the
first [Fe/H] values in the literature. The abundance analysis, based on 
Kurucz model atmospheres and LTE assumption, was computed in the canonical way.
Besides [Fe/H] values, the physical parameters $T_{\rm eff}$, $\log g$ (corrected for
acceleration of the atmosphere), microturbulent velocity, and upper limit for the
projected rotational velocity $\langle v_e \sin i \rangle$ were also determined 
from the spectra. For a few stars in common, our new abundances are in good agreement
with the results of other recent abundance analyses in the literature. 

The observed period ratios of the program stars show strong dependence on
the iron abundance and only weak dependence on the fundamental period. This may
suggest that the scatter of the known Cepheids in the Petersen diagram
 ($P_1/P_0$ vs. $\log P_0$) is mainly due to differences in their metallicity.
 

\begin{acknowledgements}
This work was supported by Hungarian OTKA Grants No. TS 049872, T 042509 and T 046207. 
EP and LS acknowledge financial support in the framework of the Italian-Hungarian T\'eT cooperation 
(Project I-24/1999). An anonymous referee has provided useful comments and suggestions, which
is also acknowledged here.  The authors also thank Monica Rainer for checking the normalized and calibrated spectra.
Thanks are also due to ESO and DDO for spectroscopic observations in La Silla, Chile and in Canada.
The NASA ADS Abstract Service was used  to access data and references.
\end{acknowledgements}

\end{document}